\documentstyle[12pt]{article}
\newcounter{equanumber}
\def\eq{%
\addtocounter{equanumber}{1}
\rm (\theequanumber )
}
\newcounter{EQA}\newcounter{EQB}\newcounter{EQC}
\newcounter{EQD}\newcounter{EQE}\newcounter{EQF}
\newcounter{EQG}
\def\Set#1{\addtocounter{#1}{\theequanumber}}
\begin{document}
\baselineskip=24truept
\thispagestyle{empty} \noindent
\rightline{HUEAP-007}
\rightline{May 19, 1994}
\center{To appeare in {\bf Nuovo Cimento}}
\begin{center}\begin{large}\begin{bf}
\vspace{1.5cm} Does gravitational
wave propagate in the five dimensional space-time with Kaluza-Klein
monopole?
\end{bf}\end{large}\end{center}
\vspace{.7cm}
\begin{center}
Osamu ABE\\
\vspace{.7cm}
\begin{it}
Physics Laboratory, Hokkaido University of Education\\
Asahikawa Campus\\
9 Hokumoncho, Asahikawa 070, Japan\\
\end{it}
\begin{bf}
Internet: osamu@atson.asa.hokkyodai.ac.jp
\end{bf}
\end{center}
\vspace{1.0cm}
\centerline{\bf Abstract}
\begin{quotation}
\noindent The behavior of small perturbations around the Kaluza-Klein
monopole in the five dimensional space-time is investigated. The fact
that the odd parity gravitational wave does not propagate in the five
dimensional space-time with Kaluza-Klein monopole is found provided
that the gravitational wave is constant in the fifth direction.
\end{quotation}
\vspace{1.5cm}
\leftline{PACS numbers: 03.40.Kf, 04.30.+x, 11.10.Kk}
\vfill\eject
%
%
\def\laplace{{\sqcap\!\!\!\!\sqcup}}
\leftline{\bf Introduction}
\vspace{.5cm}
Kaluza and Klein\cite{Kaluza} showed an elegant way of unifying
gravity and electromagnetism. In this theory, they are unified as a pure
gravity in the five dimensional manifold which is composed of
four dimensional Minkowsky space-time $M_4$ and an extra dimensional space
$S_1$. Note that the masses of massive modes in the Kaluza-Klein(K-K) theory
are tightly related with the radius of $S_1$.
The idea has been generalized \cite{deWitt}
to include non-abelian gauge theories.
Gross and Perry\cite{Gross} and independently Sorkin\cite{Sorkin} found
the regular, static and topologically stable magnetic monopoles
solution of the five dimensional Einstein equation. Iwazaki\cite{Iwazaki}
 discussed the quantum effects of the magnetic monopoles.
The discovery of
the magnetic monopole strongly indicates the unification of the fundamental
interactions.
Sundaresan and Tanaka \cite{Sundaresan}
discussed whether the space-time with Kaluza-Klein(K-K) monopole is stable
against small perturbations.

In the previous paper\cite{Abe}, we discussed the propagation of the small
perturbations around K-K metric without monopole in detail.
In this paper, we will investigate the
propagation of the small perturbations around the K-K monopole solution.

We consider K-K theory with the Einstein-Hilbert action in five dimensions
given by
$$ S={-1\over 16\pi G_5}\int d^5x\sqrt{|g_5|}R_5, \leqno\eq$$
where $G_5$ is the five dimensional gravitational constant, $g_5$ is
the determinant of five dimensional metric $g_{AB}$, and $R_5$ denotes
the five dimensional curvature scalar of the space-time.

The magnetic monopole solution is the solution of the Einstein equation
in the empty space-time
$$ R_{AB} = 0. \leqno\eq$$
We take the metric $g_{AB}$ as
$$ g_{AB}=\left(\matrix{ 1& 0& 0& 0& 0\cr
0& -V^{-1}& 0& 0& 0\cr
0& 0& -r^2V^{-1}& 0& 0\cr
0& 0& 0& -{r^2s^2+A^2V^2 \over V}& -AV\cr
0& 0& 0& -AV& -V\cr
}\right) \leqno\eq$$
\Set{EQA}
Here,
$$ V^{-1}(r) = 1 + {4M\over r},\; s = \sin\theta . \leqno\eq$$
The gauge field, $A_\mu$, is that of monopole located at the origin
$$ A_\mu (x) = \Bigl(A_t, A_r, A_\theta , A_\varphi=A)
=(0,0,0,4M\alpha (\theta ) \Bigr), \leqno\eq$$
where
$$ \alpha(\theta ) =\cases{ 1-\cos\theta&in $R_a:\{0\le
r,0\le\theta<{\pi\over 2}+\delta , 0\le\varphi <2\pi\}$\cr
-1-\cos\theta&in $R_b:\{0\le r,{\pi\over 2}-\delta <\theta\le\pi,
0\le\varphi <2\pi\}$\cr},
\leqno\eq$$
and $A_\mu$ satisfies
$$ {\bf B} ={\bf \nabla}\times {\bf A} = {4M{\bf r}\over r^3}.
\leqno\eq$$
In order to avoid the so-called NUT singularity, $M$ should satisfy\cite{Gross}
$$ M = {\sqrt{\pi G}\over 2e}, \leqno\eq$$
where $G$ is the four dimensional gravitational constant and $e$
denotes the unit electric charge. Then the magnetic charge $g$ of the
monopole is given by\cite{Gross}
$$ g = {4M\over \sqrt{16\pi G}} ={1\over 2e}. \leqno\eq$$
Thus the monopole has one unit of Dirac charge.

Now, we consider small perturbations around the monopole solution. Then
the total metric $g^T_{AB}$ is given by
$$ g^T_{AB} = g_{AB} + h_{AB}. \leqno\eq$$
According to the approach of Regge and Wheeler\cite{Regge}, we can
divide small perturbation into even and odd parity parts:
$$ h_{AB}^{even}(t,r,\theta ,\varphi , x^5) =\sum_{n^5=-\infty}^{\infty}
\leqno\eq$$
$$ \left(\matrix{ H_0 & H_1 & H_3 \nabla_2 & H_3 \nabla_3 &
H_{0p} \cr
Sym & H_2 & H_4 \nabla_2 & H_4 \nabla_3 & H_{1p} \cr
Sym & Sym & r^2\gamma_{22}K +L \nabla_2\nabla_2 & L
\nabla_2\nabla_3 & H_5 \nabla_2\cr
Sym & Sym & Sym & r^2\gamma_{33}K +L \nabla_3\nabla_3 & H_5
\nabla_3 \cr
Sym & Sym & Sym & Sym & H_{2p} \cr
}\right) $$
$$ \times Y_{q\ell m}(\theta ,\varphi ) e^{-i\omega t}e^{in^5x^5/R}$$
and
$$ h_{AB}^{odd}(t,r,\theta ,\varphi ,x^5) =\sum_{n^5=-\infty}^{\infty}
\leqno\eq$$
$$ \left(\matrix{ 0& 0& {\epsilon_2}^3h_0 \nabla_3&
{\epsilon_3}^2h_0 \nabla_2 & 0\cr
Sym& 0& {\epsilon_2}^3h_1 \nabla_3& {\epsilon_3}^2h_1 \nabla_2 &
0\cr
Sym& Sym& 2{\epsilon_2}^3h_2 \nabla_3\nabla_2& h_2
[{\epsilon_2}^3\nabla_3\nabla_3+{\epsilon_3}^2\nabla_2\nabla_2]&
{\epsilon_2}^3h_3 \nabla_3\cr
Sym& Sym& Sym& 2{\epsilon_3}^2h_2 \nabla_2\nabla_3&
{\epsilon_3}^2h_3 \nabla_2\cr
Sym& Sym& Sym& Sym& 0\cr
}\right) $$
$$ \times Y_{q\ell m}(\theta ,\varphi ) e^{-i\omega t}e^{in^5x^5/R}.$$
Here, $R$ is the radius of the circle in the fifth dimension and we
have used
$$ \gamma_{AB} = {g_{AB}\over r^2}, \leqno\eq$$
$$ {\epsilon_2}^3 \equiv g^{3A}\sqrt{|g_5|}e_{012A5} = -{1\over s},
\leqno\eq$$
and
$$ {\epsilon_3}^2 \equiv g^{2A}\sqrt{|g_5|}e_{013A5} = s, \leqno\eq$$
where $e_{ABCDE}$'s are totally antisymmetric Levi-Chivita symbols in
five dimensions. In Eqs.(11) and (12), $Y_{q\ell m}$ denotes the
monopole harmonics\cite{Wu}, which is introduced so as to avoid the
singularity at $\theta =\pi$ and which satisfies following eigen value
equations.
$$ {\bf L}^2Y_{q\ell m} \equiv\Bigl\{-{1\over s}\partial_\theta
s\partial_\theta
-{1\over s^2}\partial_\varphi^2-{2q\alpha
(\theta ) L_z\over s^2}\Bigr\}Y_{q\ell m} =\ell (\ell +1) Y_{q\ell
m},\leqno\eq$$
$$ L_zY_{q\ell m}\equiv\Bigl\{ -i\partial_\varphi
-q\alpha (\theta ) -q\cos\theta\Bigr\} Y_{q\ell m}=mY_{q\ell m}.$$
The monopole harmonics $Y_{q\ell m}$ coincides with the usual spherical
harmonics $Y_{\ell m}$ when $q=0$. The number $q$ is defined by $q=eg$
and which equals to ${1\over 2}$ in the case of K-K monopole.

We will consider small perturbations with $n^5=0$ that is we consider
only perturbations which are constant in the fifth direction or
equivalently we consider massless mode only. We impose
the gauge conditions on the small perturbations as follows
$$ \nabla_Bh^{AB} = 0, \leqno\eq$$
where $h^{AB}=-g^{AC}g^{BD}h_{CD}$. We derive equations for small
perturbation $h$ as follows. The Ricci tensor will be $R_{AB}$ if it is
calculated from $g_{AB}$ and $R_{AB}+\delta R_{AB}$ if it is calculated
{}from $g_{AB}+h_{AB}$. We can get second order equations on the
perturbation $h$ if we impose the condition $\delta R_{AB}=0$. This
condition implies that the perturbed space is also empty space. The
explicit form of the equations are
$$ \delta R_{AB} = {{{R_A}^C}_B}{^D}h_{CD} + {\laplace\over 2}h_{AB} =
0. \leqno\eq$$
\vfill\eject
\leftline{\bf Propagation of odd perturbations}
\vspace{.5cm}
We have studied the gravitational wave propagation in the K-K vacuum
without magnetic monopole\cite{Abe}. In this paper, we investigate whether
the odd perturbations can propagate in the five dimensional space-time
with K-K monopole provided that the perturbations are constant in the
fifth direction.

The transversality conditions become
$$ \nabla_Ah^{0A}_{odd}=0\, ({\rm automatically \, satisfied}),
\leqno\eq$$
$$ \nabla_Ah^{1A}_{odd}={e^{-i\omega t}AV^2(1+V) Y_{q\ell m}^{(1,0)
}h_3\over sr^3}=0, \leqno\eq $$
\Set{EQB}
$$ \nabla_Ah^{2A}_{odd}={e^{-i\omega t}V\over s^3r^4}\Bigl[
h_3\Bigl\{2AcsVY_{q\ell m}^{(1,0) }+AVY_{q\ell m}^{(0,2) }+s^3r(V-1)
Y_{q\ell m}^{(1,0) }\Bigr\}\leqno\eq$$
$$ +h_2\Bigl\{-Vs^2\triangle_{\theta\varphi}Y_{q\ell m}^{(0,1) }
+ s^2V(V^2-2V-1) Y_{q\ell m}^{(0,1) } \Bigr\}$$
$$ -s^2VY_{q\ell m}^{(0,1) }(r^2h_1) '-i\omega h_0s^2r^2Y_{q\ell
m}^{(0,1) } \Bigr]=0,$$
$$ \nabla_Ah^{3A}_{odd}={e^{-i\omega t}V\over s^4r^5} \Bigl[
h_2V\Bigl\{AV(V-1) s^2\triangle_{\theta\varphi}Y_{q\ell m}
+ \leqno\eq$$
$$ r\Bigl(-c^2sY_{q\ell m}^{(1,0) }-2cY_{q\ell m}^{(0,2) }+cs^2Y_{q\ell
m}^{(2,0) }+sY_{q\ell m}^{(1,2) }+s^3Y_{q\ell m}^{(3,0) }
-s^3V^2Y_{q\ell m}^{(1,0) }+2s^3VY_{q\ell m}^{(1,0) }\Bigr) \Bigr\}$$
$$ -AsrVh_3Y_{q\ell m}^{(1,1) }+is^3\omega r^3h_0Y_{q\ell m}^{(1,0)
}+s^3rVY_{q\ell m}^{(1,0) }(r^2h_1) ' \Bigr] =0$$
and
$$ \nabla_Ah^{5A}_{odd}=-{e^{-i\omega t}AV\over s^4r^5}\Bigl[
h_2V\Bigl\{AV(V-1) s^2\triangle_{\theta\varphi}Y_{q\ell m}\leqno\eq$$
$$ +r\Bigl(-c^2sY_{q\ell m}^{(1,0) }-2cY_{q\ell m}^{(0,2)
}+cs^2Y_{q\ell m}^{(2,0) }+sY_{q\ell m}^{(1,2) } +s^3Y_{q\ell m}^{(3,0)
}-s^3V^2Y_{q\ell m}^{(1,0) }+2s^3VY_{q\ell m}^{(1,0) }\Bigr) \Bigr\}$$
$$ -AsrVh_3Y_{q\ell m}^{(1,1) }+is^3\omega r^3h_0Y_{q\ell m}^{(1,0) }
+s^3rVY_{q\ell m}^{(1,0) }(r^2h_1) ' \Bigr]=0$$
Here, we have used
$$ \triangle_{\theta\varphi} = s^{-1}\partial_\theta
s\partial_\theta
+ s^{-2}\partial_\varphi^2,
\leqno\eq$$
and
$$Y^{(j,k)}_{q\ell m}=\partial_\theta^j\partial_\varphi^k Y_{q\ell m}.
\leqno\eq$$
Further, wave equations for small perturbation with $B=5$ are given by
$$ 2\delta R_{05}^{odd}=-{e^{-i\omega t}V^2(V-1)
h_0\triangle_{\theta\varphi} Y_{q\ell m}\over r^3}=0, \leqno\eq$$
\Set{EQC}
$$ 2\delta R_{15}^{odd}=-{e^{-i\omega t}V^2(V-1)
h_1\triangle_{\theta\varphi} Y_{q\ell m}\over r^3}=0, \leqno\eq$$
\Set{EQD}
$$ 2\delta R_{25}^{odd}=-{e^{-i\omega t}\over s^3r^4}\Bigl[
V^2h_2\Bigl\{ As^2V(V-1) ^2\triangle_{\theta\varphi}Y_{q\ell m}
+s^3r(V-1) \triangle_{\theta\varphi}Y_{q\ell m}^{(1,0) }\leqno\eq$$
$$ -cr(V-1) \Bigl(csY_{q\ell m}^{(1,0) }+2Y_{q\ell m}^{(0,2) }\Bigr)
-s^3rV(V-1) (V-2) Y_{q\ell m}^{(1,0) } \Bigr\}$$
$$ +srh_3\Bigl\{ AV^2(V-1) Y_{q\ell m}^{(1,1)
}-sr\Bigl(V\triangle_{\theta\varphi} -2V^2(V-1) +r^2\omega ^2\Bigr)
Y_{q\ell m}^{(0,1) }\Bigr\}$$
$$ -s^2r^4VY_{q\ell m}^{(0,1) }h_3'' +s^3r^2V^2(2V+1) (V-1) Y_{q\ell
m}^{(1,0) }h_1 \Bigr]=0, $$
$$ 2\delta R_{35}^{odd}={e^{-i\omega t}\over s^2r^4}\Bigl[
h_3\Bigl\{s\Bigl(A^2V^3(3V-1) (V-1) -2AcrV^2(V-1) +c^2r^2V\leqno\eq$$
\Set{EQE}
$$ +s^2r^2V^3-s^2r^4\omega ^2\Bigr) Y_{q\ell m}^{(1,0) }
-s^3r^2V\triangle_{\theta\varphi}Y_{q\ell m}^{(1,0) } -rV\Bigl(AV(V-1)
-2cr\Bigr) Y_{q\ell m}^{(0,2) } \Bigr\}$$
$$ -s^3r^4VY_{q\ell m}^{(1,0) }h_3'' +s^2rV^2h_2\Bigl\{ -(V-1)
\triangle_{\theta\varphi}Y_{q\ell m}^{(0,1) } +(V-1) (V^2-2V-1)
Y_{q\ell m}^{(0,1) }\Bigr\}$$
$$ -s^2r^2V^2(2V+1) (V-1) Y_{q\ell m}^{(0,1) }h_1\Bigr]=0$$
and
$$ 2\delta R_{55}^{odd}={e^{-i\omega t}V^2(V-1) h_3\over s^2r^4}\Bigl[
AsV(3V-1) Y_{q\ell m}^{(1,0) } -2s^2r\triangle_{\theta\varphi}Y_{q\ell
m}\Bigr]=0 \leqno\eq$$

Eq.(\theEQB) implies
$$ h_3 = 0, \leqno\eq$$
\Set{EQF}
because $Y_{q\ell m}^{(1,0) }$ with $q\neq 0$ does not vanish
identically. Eq.(\theEQC) implies
$$ h_0 = 0. \leqno\eq$$
{}From Eq.(\theEQD), we find
$$ h_1 = 0. \leqno\eq$$
\Set{EQG}
Eqs.(\theEQE), (\theEQF) and (\theEQG) tell us
$$ h_2 = 0. \leqno\eq$$
Thus, we find that small perturbations with odd parity are forbidden,
provided that the perturbations are constant in the fifth direction.

\vspace{.5cm}
\leftline{\bf Summary and conclusion}
\vspace{.5cm}
In this paper, we have studied the properties of the small
perturbations around the five dimensional Kaluza-Klein metric with
magnetic monopole. We found that
in the Kaluza-Klein space-time with monopole, there
was no odd parity solution provided that the solution is independent of
the fifth coordinate. The non-existence of the time dependent solution
implies the classical stability of the Kaluza-Klein monopole.

One may expect that far away from the magnetic monopole, for example,
the space time is asymptotically approximated by the usual vacuum and
all the vacuum perturbation must be allowed. But this expectation is
not true, because K-K monopole metric, which is given in Eq.(\theEQA),
does not approaches to the vacuum metric even in the space far away
{}from the origin. The existence of the K-K monopole gives
non-local intervention.

Our arguments were restricted to the five dimensions, but our arguments
would be applicable to more realistic theories.

We have discussed only the odd parity perturbations when the K-K
monopole have been taken into account. It would be of interest to study
the even parity perturbations. We hope to return to this problem in the
future.

\vspace{1cm}
\centerline{\bf Acknowledgement}
The author acknowledges useful discussions with K.
Tanaka and M. K. Sundaresan. He would like to thank Department of Physics,
the Ohio State University for their hospitality, where this
work was started. The algebraic calculations were done by the use of
REDUCE and {\it Mathematica}.
\vfill\eject

\end{document}